	\documentclass[aps,prl,twocolumn,footinbib,superscriptaddress]{revtex4-2} 
	
	\usepackage{amsmath}
	\usepackage{graphicx} 
	\usepackage[breaklinks=true,colorlinks,citecolor=blue,linkcolor=blue,urlcolor=blue]{hyperref}
	\usepackage{siunitx}
	\usepackage{soul}
	\usepackage{bm}
	\usepackage[normalem]{ulem}
	\usepackage{enumerate}  
	\usepackage{float}
	\usepackage{amssymb}
	\usepackage{mathrsfs}

\begin{document}

\title{\textcolor{black}{Experimental Demonstration of Controllable $\mathcal{PT}$ and anti-$\mathcal{PT}$ Coupling in a non-Hermitian Metamaterial}}

\author{Chang~Li$^{}$}
\email{lichangphy@gmail.com}
%\thanks{These authors contributed equally to this work.}
\affiliation{Key Laboratory of Light Field Manipulation and Information Acquisition Ministry of Industry and Information Technology and School of Physical Science and Technology Northwestern Polytechnical University Xi’an 710129, China}
\affiliation{European Center for Quantum Sciences (CESQ-ISIS, UMR7006), University of Strasbourg and CNRS, Strasbourg, France}

\author{Ruisheng~Yang$^{}$}
%\thanks{These authors contributed equally to this work.}
\affiliation{Key Laboratory of Light Field Manipulation and Information Acquisition Ministry of Industry and Information Technology and School of Physical Science and Technology Northwestern Polytechnical University Xi’an 710129, China}
\affiliation{Institute of Precision Optical Engineering, School of Physics Science and Engineering, Tongji University, Shanghai 200092, China}
\affiliation{Shanghai Frontiers Science Research Base of Digital Optics, Tongji University, Shanghai 200092, China}

\author{Xinchao~Huang$^{}$}
\affiliation{Key Laboratory of Light Field Manipulation and Information Acquisition Ministry of Industry and Information Technology and School of Physical Science and Technology Northwestern Polytechnical University Xi’an 710129, China}
\affiliation{European XFEL GmbH, Holzkoppel 4, 22869 Schenefeld, Germany.}

\author{Quanhong~Fu$^{}$}
\affiliation{Key Laboratory of Light Field Manipulation and Information Acquisition Ministry of Industry and Information Technology and School of Physical Science and Technology Northwestern Polytechnical University Xi’an 710129, China}

\author{Yuancheng~Fan$^{}$}
\email{phyfan@nwpu.edu.cn}
\affiliation{Key Laboratory of Light Field Manipulation and Information Acquisition Ministry of Industry and Information Technology and School of Physical Science and Technology Northwestern Polytechnical University Xi’an 710129, China}

\author{Fuli~Zhang$^{}$}
\email{fuli.zhang@nwpu.edu.cn}
\affiliation{Key Laboratory of Light Field Manipulation and Information Acquisition Ministry of Industry and Information Technology and School of Physical Science and Technology Northwestern Polytechnical University Xi’an 710129, China}

\begin{abstract}

Non-Hermiticity has recently emerged as a rapidly developing field due to its exotic characteristics related to open systems, where the dissipation plays a critical role. In the presence of balanced energy gain and loss with environment, the system exhibits parity-time ($\mathcal{PT}$) symmetry, meanwhile as the conjugate counterpart, anti-$\mathcal{PT}$ symmetry can be achieved with dissipative coupling within the system. Here, we demonstrate the coherence of complex dissipative coupling can control the transition between $\mathcal{PT}$ and anti-$\mathcal{PT}$ symmetry in an electromagnetic metamaterial. Notably, the achievement of the anti-$\mathcal{PT}$ symmetric phase is independent of variations in dissipation.  Furthermore, we observe phase transitions as the system crosses exceptional points in both anti-$\mathcal{PT}$ and $\mathcal{PT}$ symmetric metamaterial configurations, achieved by manipulating the frequency and dissipation of resonators. This work provides a promising metamaterial design for broader exploration of non-Hermitian physics and practical application with controllable Hamiltonian. 

\end{abstract}

\maketitle

\textit{Introduction}\textbf{---}
In a realistic physical system, there always exists energy exchange with the outside environment and  complex coupling between inside components. To study such open systems, non-Hermiticity has emerged as a rapidly developing direction, where the Hamiltonian possesses non-orthogonal eigen-modes with complex spectra~\cite{PhysRevLett.80.5243, Bender_2007, NonHermiYuto, RevModPhys.93.015005}. In particular, non-Hermiticity with parity-time ($\mathcal{PT}$) symmetry, arising from a balance of energy gain and loss, has been achieved in various physical systems, including optical systems~\cite{el2007theory, PhysRevLett.103.093902, ruter2010observation, regensburger2012parity,feng2013experimental, hodaei2014parity, peng2014parity, feng2014single, chang2014parity,ozdemir2019parity}, optomechanics~\cite{PhysRevLett.113.053604, xu2016topological}, acoustics~\cite{shi2016accessing, PhysRevX.4.031042, fleury2015invisible,huang2024acoustic, PhysRevX.6.021007}, LRC circuits~\cite{PhysRevLett.110.234101, assawaworrarit2017robust, PhysRevLett.123.213901, PhysRevLett.128.065701}, and atomic systems~\cite{li2019observation, PhysRevLett.126.083604, wu2019observation, PhysRevLett.127.090501}.
When adjusting the energy exchange with environment, the exceptional point can be observed in a $\mathcal{PT}$ symmetric system, where the eigen-modes coalesce with the same eigen-energy value. Upon crossing an exceptional point, the system can transition from a symmetric phase to a symmetry-broken phase, where eigen-energies shift from real to imaginary values~\cite{Heiss_2012,PhysRevLett.103.093902}.

As the counterpart, anti-$\mathcal{PT}$ symmetric non-Hermiticity has also been explored recently~\cite{peng2016anti, PhysRevLett.123.193604,yang2022radiative, PhysRevLett.130.213603,bergman2021observation,choi2018observation, guo2023level,PhysRevLett.124.053901,li2019anti,PhysRevLett.131.066601,arwas2022anyonic,PhysRevLett.125.147202}, and unique properties have been demonstrated such as unit refraction~\cite{peng2016anti}, coherent switch~\cite{PhysRevLett.120.123902} and energy-difference conserving dynamics~\cite{choi2018observation}. The origin of anti-commutation between the anti-$\mathcal{PT}$- symmetric non-Hermitian Hamiltonian and parity-time reversal operators is the dissipative coupling within the system. The coupling term is purely imaginary and anti-conjugate between components in the anti-$\mathcal{PT}$ symmetric system~\cite{PhysRevA.96.053845}; the coupling between components $i, j$ has the relation: $\kappa_{ij}=-\kappa_{ji}^*$, which is distinguished from normal complex conjugate coupling, $\kappa_{ij}=\kappa_{ji}^*$. Therefore, we aim to investigate whether this unique dissipative coupling can be coherently controlled, similar to near-field coupling between resonators, to broadly explore the non-Hermiticity. For instance, the dissipative coupling can be modulated from imaginary to real value if its phase can be coherently controlled, and further accomplishing the transition from anti-$\mathcal{PT}$ symmetry to $\mathcal{PT}$ symmetry.

%%%%%%%%%%%%%%%%%% Fig 1 Setting %%%%%%%%%%%%%%%%%%%%%%%
\begin{figure*}[!tbp]
  \centering
  \includegraphics[width=14cm]{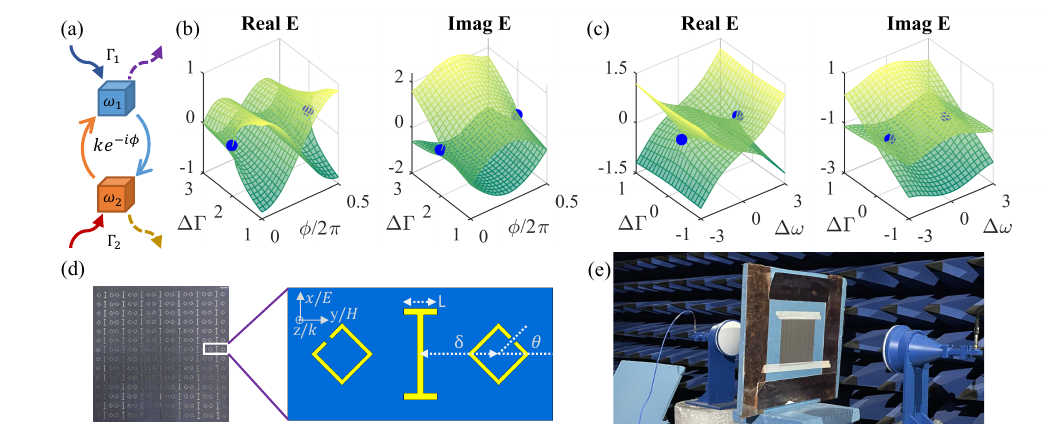}\\
  \caption{\textbf{Non-Hermiticity in metamaterials} \textbf{(a)} Diagram of two strongly coupled resonators. 
The eigen-frequencies and dissipation rates are denoted by $\omega_{i}$ and $\Gamma_{i}~(i=1,~2)$. The coupling strength between them is $k$ with phase delay $\phi$. \textbf{(b)} Eigen-energy as a function of $\phi$ and $\Delta\Gamma$ in Eq.~(3). The real and imaginary parts of $E_{\pm}$ are calculated with fixed $k=1$ and $\Gamma_1=1$. The blue dots denote exceptional points where two eigen-modes degenerate with same real and imaginary parts. \textbf{(c)} Eigen-energy as a function of $\Delta\omega$ and $\Delta\Gamma$ in Eq.~(4) with fixed $k=1$ and $\Gamma_1=1$. The blue dots are exceptional points. \textbf{(d)} Metamaterial configuration. The image to the left shows one sample used for experiments. All samples contain $7\times14$ units and each unit is $10\times20$ mm$^2$. The sketch of one unit is shown on the right. The I-shape cut wire with short arm length $L$ is at the center, and a couple of SRR with rotation angle $\theta$ are symmetrically located on both sides, with a distance $\delta$ to the center. The probing EM field is alongside z-axis and perpendicular to the sample, and the electric component is parallel to the long-arm of cut wire alongside x-axis. \textbf{(e)} Image of the experiment setup. Linear-polarized microwave is generated from the first horn and collimated by a lens with a focus of $1.2$ m. The sample mount is located at the focus, which is a flat microwave absorber with a window at center, and samples are stuck to the surface.}
	\label{fig1}
\end{figure*}
%%%%%%%%%%%%%%%%%%%%%%%%%%%%%%%%%%%%%%%%%%%%%%%%%

In this work, we demonstrate such complex coupling can be achieved and further facilitate the transition between anti-$\mathcal{PT}$ and $\mathcal{PT}$ symmetry in an electromagnetic metamaterial. The coupling phase is modulated by varying the propagation distance between the resonators. Achieving anti-$\mathcal{PT}$ symmetry is accomplished by adjusting coupling phase to $\pi/2$, where the real part of eigen-energies coalesces. Notably, the anti-$\mathcal{PT}$ symmetric phase is shown to be independent of dissipation, which distinguishes it from $\mathcal{PT}$ symmetric phase. Furthermore, we observe phase transitions in both anti-$\mathcal{PT}$ and $\mathcal{PT}$ symmetric metamaterial configurations as they cross exceptional points with controllable resonant frequency and dissipation rate. All results are theoretically analyzed and further validated through both experiments and simulations. Our work bridges two critical aspects of non-Hermitian physics, and provides a practical metamaterial design for controlling the non-Hermitian Hamiltonian.

\textit{Non-Hermitian in two resonators}\textbf{---}Our model is composed of two strong-coupled resonators, as shown in Fig.~\ref{fig1}(a). Two resonators are coupled to each other, with additional phase delay $\phi$ when interaction is propagated by, for example, photons~\cite{wang2020electromagnetically} or flying atoms~\cite{peng2016anti}. The coupling efficiency \textbf{$\kappa$} is denoted as a complex number \textbf{$\kappa$}$=ke^{-i\phi}$, where $k$ is a real positive number standing for the coupling strength. Both resonators can be excited by external field and have individual energy losses, with resonant frequencies and dissipation rates denoted as $\omega_{i}$ and $\Gamma_{i}~(i=1,~2)$ respectively. The dynamics of the resonators can be described by: 
\begin{equation}
\label{eq1}
\begin{split}
& i\frac{da_1}{dt} =\omega_1a_1-i\Gamma_1a_1+\kappa a_2 \\
& i\frac{da_2}{dt} =\omega_2a_2-i\Gamma_2a_2+\kappa a_1
\end{split}
\end{equation}
where the $a_i~(i=1, 2)$ correspond to the resonating amplitudes. The effective Hamiltonian, $\hat{H}$, can be extracted and further decomposed into Hermitian and anti-Hermitian parts: 
\begin{equation}
\label{eq2}
\begin{aligned}
\hat{H}&=
\left[
\begin{matrix}
&\omega_1-i\Gamma_1 & \kappa \\\
&\kappa & \omega_2-i\Gamma_2
\end{matrix}
\right] \\
&=\left[
\begin{matrix}
&\omega_1 & Re(\kappa) \\\
&Re(\kappa) & \omega_2
\end{matrix}
\right]-i\left[
\begin{matrix}
&\Gamma_1 & -Im(\kappa) \\\
&-Im(\kappa) & \Gamma_2
\end{matrix}
\right]
\end{aligned}
\end{equation}
From Eq.~\ref{eq2}, the non-Hermitian clearly originates from the dissipation of resonators and the complex coupling between them. In the following, we check the impact of these parameters on the eigen-energies of $\hat{H}$. 

If we consider $\omega_1=\omega_2=0$, the eigen-energies are: 
\begin{equation}
\label{eq3}
\begin{aligned}
E_{\pm}=-i\bar\Gamma\pm \sqrt{k^2e^{-2i\phi}-{\Delta\Gamma}^2/4},
\end{aligned}
\end{equation}
where $\bar\Gamma=(\Gamma_1+\Gamma_2)/2$ and $\Delta\Gamma=\Gamma_2-\Gamma_1$. In Fig.~\ref{fig1}(b), we provide one example of the eigen-energies while varying $\phi$ and $\Delta\Gamma$, where the real and imaginary parts are shown separately and correspond to the resonant frequency and loss of the eigen-modes respectively. When the coupling phase $\phi=0$ (or $\pi$), $E_{\pm}$ will be in the typical form from a system with gain-loss type $\mathcal{PT}$symmetric non-Hermitian. The exceptional points are at $\Delta\Gamma=\pm2k$ (highlighted in blue dots), and we can observe the transition between $\mathcal{PT}$ symmetric phase and symmetry broken phase crossing the exceptional points. On the other hand, the anti-$\mathcal{PT}$ symmetry can be achieved when coupling efficiency is purely imaginary with $\phi=\pi/2$. In this case, the real parts of the eigen-energies $E_{\pm}$ coincide, and the system is in anti-$\mathcal{PT}$ symmetric phase. This phase is independent of dissipation and the coincidence of real energies is preserved with different $\Delta\Gamma$, which distinguishes itself with $\mathcal{PT}$ symmetric phase. 
%%%%%%%%%%%%%%%%%% Fig 2 Setting %%%%%%%%%%%%%%%%%%%%%%%
\begin{figure}[tb]
  \centering
  \includegraphics[width=8cm]{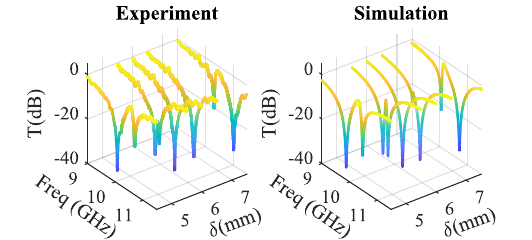}\\
  \caption{\textbf{Achieve anti-$\mathcal{PT}$ symmetry with complex coupling.} The transmission spectra are obtained both from experiment (left) and simulation (right) to find out the anti-$\mathcal{PT}$ symmetry. Five groups of spectra correspond to the distance $\delta=[4.5,~5.5,~6.0,~6.5,~7.5]$~mm. Other configuration parameters are fixed, $L=2.5$~mm and $\theta=0^{\circ}$. The transparent window vanishes at $\delta=6$~mm which indicates anti-$\mathcal{PT}$ symmetry with coupling phase $\phi=\pi/2$}
	\label{fig2}
\end{figure}
%%%%%%%%%%%%%%%%%%%%%%%%%%%%%%%%%%%%%%%%%%%%%%%%%

To observe transition between anti-$\mathcal{PT}$ symmetric phase and symmetry broken phase, the eigen-frequency of resonators should be modified correspondingly. Considering the coupling phase $\phi=\pi/2$ and resonant frequencies $\omega_1=-\omega_2=\Delta\omega/2$, the eigen-energies are: 
\begin{equation}
\label{eq4}
\begin{aligned}
E_{\pm}=-i\bar{\Gamma}\pm \sqrt{(\Delta\omega-i\Delta\Gamma)^2/4-k^2}.
\end{aligned}
\end{equation}
As shown in Fig.~\ref{fig1}(c), two exceptional points (blue dots) exist when $\Delta\Gamma=0$ and $\Delta\omega=\pm2k$. Between the exceptional points, only real part of $E_{\pm}$ is coincident, which indicates the system is in anti-$\mathcal{PT}$ symmetric phase. When $|\Delta\omega|>2k$, the imaginary parts are the same, and the system is in anti-$\mathcal{PT}$ symmetry broken phase. 

\textit{Metamaterial design}\textbf{---}To achieve non-Hermiticity described above with a metamaterial, three parameters should be controlled, the coupling phase $\phi$, the differences of resonant frequency $\Delta\omega$ and dissipation $\Delta\Gamma$. The metamaterial design is shown in Fig.~\ref{fig1}(d). It contains $14\times7$ periodic units, and each unit is composed of a I-shape cut wire resonator and a couple of split-ring resonator (SRR). Three geometry parameters are varied to control the Hamiltonian: the distance between resonators $\delta$ controls the coupling phase $\phi$ as the propagation distance; the short-arm length of the cut wire $L$ changes the resonance frequency $\omega_1$; the rotation angle of SRR $\theta$ adjusts the dissipation rate $\Gamma_2$ owing to power broadening~\cite{supm}. The transmission spectra of metamaterials are measured in the microwave shielded room, as illustrated in Fig.~\ref{fig1}(e). Two horns with microwave lenses serve as the microwave emitter and receiver, and the metamaterials are mounted on the microwave absorber at the focus of the lenses. The vector network analyzer is used as microwave source and collects the transmission spectra, which are verified with software simulation using CST Microwave Studio~\cite{supm}.

%%%%%%%%%%%%%%%%%% Fig 3 Setting %%%%%%%%%%%%%%%%%%%%%%%
\begin{figure}[!tbp]
  \centering
  \includegraphics[width=8cm]{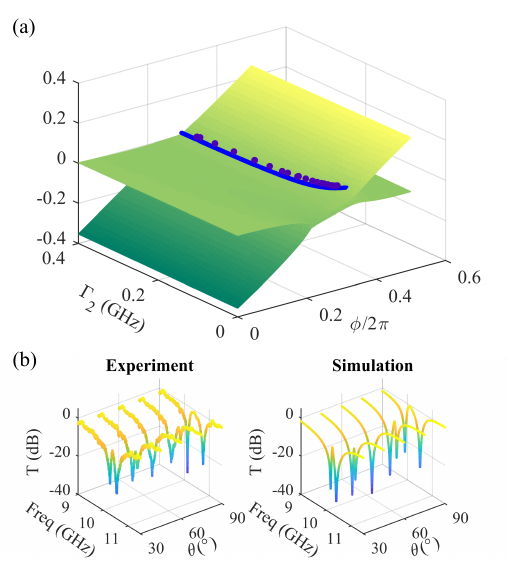}\\
  \caption{\textbf{Anti-$\mathcal{PT}$ symmetry independent of dissipation variation} \textbf{(a)} The real part of $E_{\pm}$ in Eq.~(4) as a function of $\phi$ and $\Gamma_2$. The parameters used for calculation are extracted from experimental spectra, and they are as followed: $\omega_1=0$, $\Gamma_1=0.34$~GHz, $k=30$~MHz. $\omega_2=-40$~MHz when $\phi=\pi/2$, and it varies linearly 0.6~GHz/$\pi$ with $\phi$. \textbf{(b)} Transmission spectra from experiment and simulation. Samples used here have fixed $L=2.5$~mm and $\delta=7$~mm. $\theta$ is varied from $30^{\circ}$ to $90^{\circ}$, and transmission peak vanishes at $\theta=60^{\circ}$.}
	\label{fig3}
\end{figure}
%%%%%%%%%%%%%%%%%%%%%%%%%%%%%%%%%%%%%%%%%%%%%%%%%
%%%%%%%%%%%%%%%%%% Fig 4 Setting %%%%%%%%%%%%%%%%%%%%%%%
\begin{figure*}[!tbp]
  \centering
  \includegraphics[width=18cm]{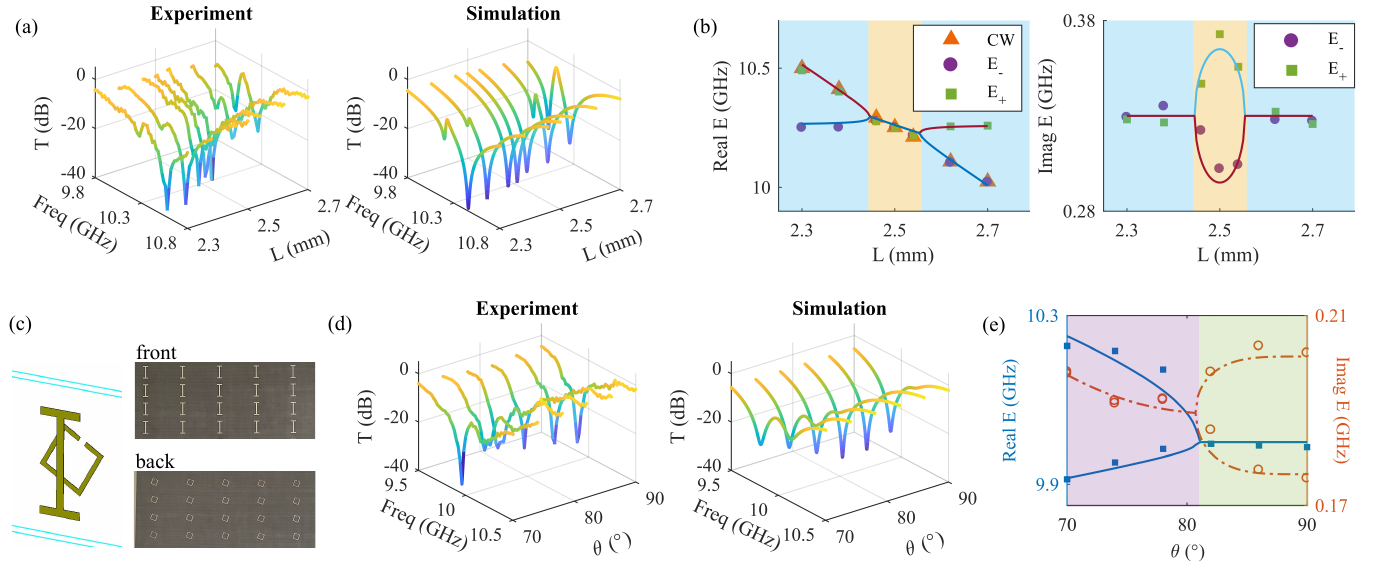}\\
  \caption{\textbf{Anti-$\mathcal{PT}$ and $\mathcal{PT}$symmetry phase transition} \textbf{(a)} Transmission spectra of anti-$\mathcal{PT}$ symmetry phase transition. The spectra is obtained with $L=[2.30,~2.38,~2.46,~2.50,~2.54,~2.62,~2.70]$~mm, with fixed $\delta=6$~mm and $\theta=0^{\circ}$. \textbf{(b)} Eigen-energy of anti-$\mathcal{PT}$ symmetry phase transition. The real (left) and imaginary (right) parts of eigen-energy are extracted from the experimental transmission spectra. The purple and green dots represent $E_{\pm}$, and red triangles are the resonant frequencies of the cut wire (CW). The coupling strength, $k$, is estimated first with the frequencies of the eigen-modes, and then the real and imaginary parts of eigen-energy are calculated afterwards as the blue and red curves showing here. \textbf{(c)} Sample for $\mathcal{PT}$ symmetry phase transition. The size of each unit is still the same, but the SRRs are at the back surface with small center offset $\delta=0.2$~mm from cut wire. we use $L=2.8$~mm to keep resonant frequencies of the cut wire and SRR are both at $10$~GHz. \textbf{(d)} Transmission spectra of $\mathcal{PT}$ symmetry phase transition. $\theta$ is linearly varied from $70^{\circ}$ to $90^{\circ}$. Two resonant modes coincide around $\theta=82^{\circ}$. \textbf{(e)} Eigen-energy of $\mathcal{PT}$ symmetry phase transition. The real (blue square) and imaginary (orange circle) parts of eigen-energy are obtained from the experiment spectra. The spectra of real (blue curve) and imaginary (orange dash dot) parts are drawn after fitting coupling strength as 151~MHz with obtained resonant frequencies. The pink (green) region represents the $\mathcal{PT}$ symmetric (symmetry broken) phase.
  } \label{fig4}
\end{figure*}
%%%%%%%%%%%%%%%%%%%%%%%%%%%%%%%%%%%%%%%%%%%%%%%%%
\textit{Achieve anti-$\mathcal{PT}$ symmetry}\textbf{---}Varying coupling phase can bring the system ranging from $\mathcal{PT}$ symmetry to anti-$\mathcal{PT}$ symmetry. Here, we first check the anti-$\mathcal{PT}$ symmetry by modifying the distance $\delta$ from $4.5$~mm to $7.5$~mm. As shown in Fig.~\ref{fig2}, a typical electromagnetically induced transparency (EIT)-like spectrum is exhibited when $\delta=4.5$~mm, with two dips corresponding to two eigen-modes. As $\delta$ is increased, the frequency difference between two modes decreases until $\delta=6$~mm, where the transparency window completely vanishes. Beyond that point, two eigen-modes reappear, indicated by the clear presence of two dips in the EIT-like spectrum. Both experiment and simulation confirm that the Hamiltonian becomes anti-$\mathcal{PT}$ symmetric when the coupling phase becomes $\pi/2$ as analyzed in Eq.~\ref{eq3}, with $\delta=6$~mm approximating one-quarter of the wavelength (around 30 mm at 10 GHz).

Based on theoretical discussion, the anti-$\mathcal{PT}$ symmetric phase can be indentified by proving its independence of dissipation variation. We verify this property by rotating SRR with angle $\theta$ to change its dissipation rate $\Gamma_2$. Initially, we calculate the real part of eigen-energies from Eq.~\ref{eq4} as shown in Fig.~\ref{fig3}(a), considering the resonant frequency drift of the SRR with different $\delta$ and $\theta$~\cite{supm}. The blue line indicates the condition to reach anti-$\mathcal{PT}$ symmetric phase, where the real parts are the same. The simulation is also performed by scanning $\delta$ and $\theta$ to find out anti-$\mathcal{PT}$ symmetry condition, as indicated by the violet dots~\cite{supm}. Both results match well with each other, showing that the anti-$\mathcal{PT}$ symmetric phase is robust against dissipation variation. Noteworthy, the phase is not always $\pi/2$ to achieve anti-$\mathcal{PT}$ symmetry; $\phi$ would be slightly larger when $\Gamma_2$ is smaller.
Hence, we verify it by scaning $\theta$ with the fixed $\delta=7$~mm ($>6~$mm). As shown in Fig~\ref{fig3}(b), the anti-$\mathcal{PT}$ symmetry happens at $\theta = 60^{ \circ}$, confirmed by both experiment and simulation. 

\textit{Phase transition}\textbf{---}With tunable resonant frequency and dissipation, we can explore the phase transition in $\mathcal{PT}$ and anti-$\mathcal{PT}$ symmetry. We start with anti-$\mathcal{PT}$ symmetric configuration with $\delta=6~$mm and $\theta = 0^{\circ}$. The resonance frequency of the cut wire is varied by changing the length, $L$. The transmission spectra are shown in Fig.~\ref{fig4}(a) from both experiment and simulation, and the real and imaginary parts of eigen-energy are obtained from frequency and width of the transmission dips, as plotted in Fig.~\ref{fig4}(b) respectively. When $L$ increases from $2.3~$mm to $2.7~$mm, the resonance frequency of the cut wire decreases from 10.5~GHz to 10~GHz (orange triangle in Fig.~\ref{fig4}(b)). From the transmission spectra, we clearly observe two resonant modes merge into one and split into two, which indicates two exceptional points are swept over and transition ranging from anti-$\mathcal{PT}$ symmetric phase to symmetry broken phase. The coupling strength is calculated, $k=32~$MHz, after fitting the real part of eigen-energies $E_{\pm}$ with experimentally obtained $\omega_i$ ~$(i=1,~2)$, and it agrees with the value fitted in Fig.~\ref{fig3}(a). In addition, the anti-$\mathcal{PT}$ symmetric phase and symmetry broken phase are acquired and shaded in yellow and blue regions respectively in Fig.~\ref{fig4}(b). 

Finally, we explore the phase transition of $\mathcal{PT}$ symmetry. The configuration of metamaterials is modified to satisfy the coupling phase $\phi=0$, which is composed of a cut wire and a SRR printed on the opposite surfaces, as shown in Fig.~\ref{fig4}(c). The angle of the SRR, $\theta$, is increased to reduce the dissipation $\Gamma_2$, and eventually to increase $\Delta\Gamma$. As shown in Fig.~\ref{fig4}(d), two eigen-modes can be clearly observed from transmission spectra when $\theta$ is small, and they start to merge till $\theta=82^{\circ}$ two modes coincide at the center frequency. Both the real and imaginary parts are extracted from transmission spectra and shown in Fig.~\ref{fig4}(e), where the $\mathcal{PT}$ symmetric region and symmetry broken region are shaded in pink and green respectively. 

\textit{Discussion}\textbf{---} In summary, the coherence of dissipative coupling is proved controlling the transition between $\mathcal{PT}$ and anti-$\mathcal{PT}$ symmetry in a non-Hermitian system. The metamaterial we design provides the flexibility to control the coupling phase, as well as the frequency and dissipation of the resonators. The anti-$\mathcal{PT}$ symmetric phase is achieved by controlling propagating distance, and proved independent of variations in dissipation, moreover, the phase transitions are observed in both anti-$\mathcal{PT}$ and $\mathcal{PT}$ symmetric metamaterial configurations. 

\textcolor{black}{
A fully tunable metamaterial can be expected to control all the parameters in non-Hermitian Hamiltonian in Eq.~\ref{eq2}, and three ingredients are required: two strong-coupled resonators, tunable eigen-frequency and dissipation of resonators, and controllable coherent dissipative coupling. Our work satisfies all these requirements but the controllability can be significantly enhanced through the application of programmable techniques in conjunction with improved metamaterial designs~\cite{supm}. The frequencies and dissipation of the resonators can be precisely manipulated by integrating controllable electronics, such as lumped circuit elements~\cite{yang2019active} and voltage-controlled graphene~\cite{balci2018electrically, ergoktas2022topological, nan2020actively, nan2023coupling}, and all the electronics components can be programmed by FPGAs (field-programmable gate arrays)~\cite{liu2022programmable, gao2023programmable, bai2022radiation}. Additionally, employing an elastic base~\cite{zhang2015mechanically} or a mechanical device~\cite{liu2019flexible} allows for effective control of the coupling phase. This enhancement affords the attainment of arbitrary non-Hermitian Hamiltonians with a programmable metamaterial, thereby facilitating further exploration on topological physics~\cite{ding2022non}. Moreover, our work unfolds various important applications in the microwave domain. For instance, sensors can be designed around either exceptional points to enhance the sensitivities to frequency perturbation in anti-$\mathcal{PT}$ symmetric configuration or the dissipation drift in $\mathcal{PT}$ symmetric configuration (see an example in~\cite{supm})~\cite{PhysRevLett.122.153902, PhysRevLett.130.227201} or to manipulate scattering with desired photon response~\cite{miri2019exceptional}; the cloak effect can be achieved by degeneracy or nonconformal distortions of non-Hermitian medium~\cite{zhang2019phase, PhysRevLett.128.183901}.} Our results can also contribute to a deeper understanding of coupling phase related collective effects, such as superradiance and subradiance~\cite{yan2023super}.

\begin{acknowledgments}
C.~Li and R.-S.~Yang contributed equally to this work. We thank Biao Yang for helpful feedback on our manuscript.
C.~Li acknowledges insightful discussions with M.-H.~Li, X.-L.~Ouyang, K.-D.~ Wang and W.-G.~Zhang,. 
This work is supported by the National Key R\&D Program of China (Grant No. 2022YFB3806000, 2023YFB3811400), the National Natural Science Foundation of China (Grant No. 12074314, 61771402, and 11674266), the Science and Technology New Star Program of Shaanxi Province (grant No. 2023KJXX-148), and the Fundamental Research Funds for the Central Universities.
\end{acknowledgments}
%\newline

%\bibliography{reference}


%\pagebreak

\end{document}